# Spheroidal Particle Stability in Semi-Solid Processing

R. A. Martinez[1], A. Karma[2], M.C. Flemings[3]

[1] Selmet, Inc., Albany, OR 97321
[2] Northeastern University, Physics Department, Boston, MA 02115
[3] Massachusetts Institute of Technology, Department of Materials Science and Engineering, Cambridge, MA 02139

## Abstract

A model for diffusion-controlled spherical particle growth is presented and solved numerically, showing how, on cooling at sufficient rate from a given fraction solid, growth velocity first increases, and then decreases rapidly when solute fields of adjacent particles overlap. An approximate analytical solution for the spherical particle growth velocity is then developed and shown to be valid until the solute fields begin to overlap. A particle stability model is next presented, building on the above analytic solution. This model permits prediction of the maximum cooling rate at which a semi-solid slurry or reheated semi-solid billet can be cooled while still retaining the spherical growth morphology. The model shows that particle stability is favored by high particle density, high fraction solid and low cooling rate. The predictions of the stability model are found to be in good quantitative agreement with experimental data collected for Al-4.5wt%Cu alloy. Engineering applications of the results obtained are discussed.

## Introduction

The desired starting material for semi-solid forming is a partially solidified (or partially melted) alloy in which the solid is present as nearly perfect spheroidal solid particles. Typically the spheroidal structure, or something approaching it, is obtained by agitation during the initial dendrite formation as the alloy is cooled through the liquidus. The agitation "breaks" the fragile dendrites, creating multiple new grains, and therefore a fine grain structure. In the usual case in practice today, the grains then grow dendritically, but will "ripen" into spheroids of greater or lesser perfection on slow cooling, holding in the liquid-solid region, or solidifying and reheating into the liquid-solid region.

It is now understood that if initial grain density is sufficiently high, growth can be perfectly spheroidal throughout the growth process, obviating the need for ripening of a dendritic structure. A concomitant advantage is that spheroids obtained in this way have none of the entrapped second phase present in most ripened dendrites.[1],[2]

From a practical point of view in semi-solid forming, the important basic issue is understanding the conditions of grain density, fraction solid and cooling rate that permit solidification in a fully non-dendritic, spheroidal mode. This work outlines those conditions; specifically comprising a modeling and experimental study of stability of the spheroidal particle interface in semi-solid Al-4.5wt%Cu alloy as a function of solid fraction and cooling rate.

**Growth Model, Numerical Solution**

A spherical particle growth model was developed for semi-solid Al-4.5wt%Cu alloy. The model assumes equilibrium at the spherical solid-liquid interface, and no solid diffusion; local solidification velocity is limited by solute diffusion away from the interface. The model is hereafter referred to as the "Liquid Diffusion Controlled," or "LDC," model. It assumes growth of a spherical particle of radius R, located at the center of a spherical volume of radius $R_T$. The value of $R_T$ represents the radius of a sphere which would occupy the same volume as will the fully solidified grain. Thus, overall solute contained within the volume element of radius $R_T$ is conserved, and the model permits determining the effect of solute field overlap on the particle growth velocity. Figure 1 shows the model schematically, illustrating the solute profile in liquid and solid, at a time when solute diffusion fields have overlapped, so that the liquid concentration at R= $R_T$ is above the initial liquid concentration, $C_0/k$. The following system of equations comprises the mathematical model:

$$\frac{\partial C}{\partial t} = D_L \left( \frac{1}{r}\frac{\partial C}{\partial r} + \frac{\partial^2 C}{\partial r^2} \right) \quad (1)$$

$$C_0(1-k)\frac{dr}{dt} = -D_L \left( \frac{\partial C}{\partial r} \right)_{r=R} \quad (2)$$

$$C_L^* - C_0 = -\frac{(T_L - T)}{m_L} + \frac{2\Gamma}{m_L}\frac{1}{R} \quad (3)$$

$$C_S^* = k C_L^* \quad (4)$$

where the variable r refers to the radial coordinate, and R is the changing particle radius. The variable T is the temperature of the system during cooling. For Al-4.5wt%Cu alloy, the constant $D_L$ is solute diffusivity in the liquid phase ($3*10^{-9}$ m$^2$ s$^{-1}$), $m_L$ is the slope of the liquidus line (-3.4 °C wt%Cu$^{-1}$), $m_S$ is the slope of the solidus line (-17.9 °C wt%Cu$^{-1}$), k is the alloy partition coefficient (0.19), $T_L$ is the liquidus temperature (649 °C), $C_0$ is the bulk solute concentration (4.5wt%Cu), $C_L^*$ is the liquid composition at the solid-liquid interface, $C_S^*$ the solute concentration of the solid at the solid-liquid interface, and $\Gamma$ is the Gibbs-Thompson coefficient ($2.4*10^{-9}$ m °C).

Equations (1) – (4) are readily solved numerically, giving the temporal dependence of the particle radius, growth velocity and the solute concentration in both the liquid and solid phases for each point contained within a sphere of radius $R_T$.

As example of the solutions obtained, Figure 2 shows liquid concentration versus radial distance, where R< r < $R_T$, and where $R_T$ = 50 μm. The example assumed the particle was allowed to grow to a 20 μm radius during slow cooling at a rate of 0.28°C s$^-$ [1], followed by a sudden increase in cooling rate to 375°C s$^{-1}$. During the initial slow cool, there is only a very small composition gradient in the liquid (i.e, solidification essentially follows the Scheil relation). However, after the rapid cooling begins (at R=20 μm) a steep solute boundary layer builds in the liquid, the boundary layer not reaching $R_T$ until R ≈ 40 μm.

Since the temperature across the volume element is uniform, the maximum solute undercooling at any time during solidification is at r=R, and is the solute difference between that at R and that $R_T$ multiplied by the liquidus slope, $m_L$: this is plotted in Figure 3. It is a maximum just as overlap of the diffusion fields occurs.

Figure 4 plots growth velocity as a function of particle radius for the same example of a particle allowed to grow to a 20 μm radius during slow cooling at a rate of 0.28°C s$^{-1}$, followed by a sudden increase in cooling rate to 375°C s$^{-1}$. When the fast cooling begins, the growth velocity first increases to almost 500 μm s$^{-1}$, and then decreases abruptly at the onset of solute field interaction, at R ≈ 40 μm. Growth velocity versus time is shown in Figure 5.

**Growth Model, Analytical Solution**

In order to develop an analytic model for interface stability in semi-solid alloys, in this section we first obtain an analytic expression for the growth velocity of a spherical particle. We consider, as in the previous section, a particle of initial radius $R_i$, present in a volume element of radius $R_T$, the liquid being essentially uniform in composition and in equilibrium with the the solid concentration at the liquid-solid interface. We then make the additional assumptions that cooling is constant and takes place sufficiently fast that the thickness of the solutal diffusion boundary layer surrounding the particle is much smaller than the particle radius, thus allowing the particle interface to be treated locally as a planar interface.

The basic equation to solve is the one-dimensional diffusion equation for the solute in the liquid

$$\frac{\partial C}{\partial t} = D_L \left( \frac{\partial^2 C}{\partial r^2} \right) \qquad (5)$$

subject to the boundary conditions at the interface

$$C_L^* - C_0 = \left( \frac{\dot{T} t}{|m_L|} \right)_{r=R} \qquad (6)$$

and

$$C_0(1-k)V = -D_L \left( \frac{\partial C}{\partial r} \right)_{r=R} \qquad (7)$$

where $\dot{T}$ is the rate of temperature change, $V$ is the particle growth velocity, and all other variables are defined in the same way as in the LDC model. The boundary condition far from the interface is:

$$C(\infty) = C_0 \qquad (8)$$

It is convenient to define the dimensionless concentration, u, as

$$u = \frac{C(r) - C_0}{C_0(1-k)}, \qquad (9)$$

and the time rate of change of the dimensionless supersaturation in the liquid, $\dot{\Omega}$, as

$$\dot{\Omega} = \frac{\dot{T}}{|m_L| C_0 (1-k)} . \qquad (10)$$

With these definitions, Equations (5) through (8) become, respectively

$$\frac{\partial u}{\partial t} = D_L \frac{\partial^2 u}{\partial r^2} \tag{11}$$

$$V = -D_L \frac{\partial u}{\partial r} \tag{12}$$

$$u(R) = \dot{\Omega} t \tag{13}$$

$$u(\infty) = 0 \tag{14}$$

A major simplification in solving this problem is that the particle radius is found to change very little during the rapid period of increase of growth velocity, as found in simulations of the LDC model. We show self-consistently below that this is a direct consequence of the fact that the particle velocity increases initially as the square root of time after the sudden change in cooling rate, and hence the particle radius increases initially slowly as the 3/2 power of time. Therefore, it is reasonable to assume that the particle radius stays constant during the initial rapid increase of the velocity. This assumption allows us to seek for a self-similar solution to Equation (11) of the form

$$u(r,t) = \dot{\Omega} t f\left(\frac{(r-R)}{(D_L t)^{1/2}}\right) \tag{15}$$

to calculate the time-dependent diffusion field around the particle of radius R. Once it is obtained, the growth velocity of the particle is found using Equation (12). To solve for the time-dependent diffusion field around the particle, Equation (15) is substituted into Equation (11) and a new scaling variable $y = \frac{(r-R)}{(D_L t)^{1/2}}$ is defined, which transforms Equation (11) into the form

$$f - \frac{y}{2}\left(\frac{df}{dy}\right) = \left(\frac{d^2 f}{dy^2}\right) \tag{16}$$

Equation (16) is solved subject to the boundary conditions f(0)=1 and f(∞)=0, which follow directly from Equations (13) and (14) together with the definition of Equation (15). The particle growth velocity is then obtained from Equation (12):

$$V = D_L \left(\frac{du}{dr}\right)_{r=R} = -\Omega(D_L t)^{1/2}\left(\frac{df}{dy}\right)_{y=0} \qquad (17)$$

Equation (17) is a second order differential equation that has a unique solution subject to the boundary conditions given by Equations (12) and (14). Solving Equation (17) numerically yields

$$-\left(\frac{df}{dy}\right)_{y=0} = 1.12 = A \qquad (18)$$

The final analytical expression relating the growth velocity of the particle subjected to a sudden increase in cooling rate is found to be

$$V = 1.12\left[\frac{\dot{T}}{|m_L|C_0(1-k)}\right](D_L \Delta t_0)^{1/2} \qquad (19)$$

where $\dot{T}$ is assumed to be constant and $\Delta t_0$ is the time starting from when the cooling rate is increased. Note that the velocity grows as the 1/2 power of this time, and hence the particle radius as the 3/2 power, as discussed previously.

To test the accuracy of Equation (19), the predicted growth velocity of a particle was compared with the growth velocity given by the LDC particle growth simulation in Figure 4 (in which $R_T = 50$ μm, $R_i = 20$ μm, and cooling rate is 376K s$^{-1}$). This is done in Figure 6. Note that Equation (19) follows the LDC simulation of the particle growth velocity very well from the time fast cooling begins to the time where solute field overlap begins (marked by the decrease in the LDC growth velocity).

Other comparisons of the analytic with the numerical model were carried out to assure agreement over the range of practical interest. Figure 7 shows one such simulation, for $R_T = 150$ μm, $R_i = 100$ μm, and cooling rate 2 °C s$^{-1}$.

## Interface Stability Model

In this section a model is developed to predict the relationship between the initial solid fraction in a semisolid alloy slurry and the maximum cooling rate at which a stable spherical morphology can be maintained. This is done by employing the basic theory of interface stability during alloy solidification developed by Mullins and Sekerka[3],[4] and later treated for a spheroidal interface (see for example Langer [5]). The treatment assumes that the solid particle is growing into an infinite liquid, an assumption that is valid in our case until diffusion fields overlap.

It was previously shown in the LDC model simulations that solute field overlap leads to a dramatic decrease in particle growth velocity. Thus, it is reasonable to assume that a particle in a slurry cooled at a given rate will always be stable if solute field overlap occurs before the interface has time to become fully unstable. The variable $\Delta t_0$ will be defined as the time it takes for solute fields to interact, and $\Delta t_u$ will be the time it takes for a particle interface to become morphologically unstable. Both $\Delta t_0$ and $\Delta t_u$ are measured from when the time the cooling rate is increased. If $\Delta t_0 < \Delta t_u$ the particle interface will always be stable, and if $\Delta t_0 > \Delta t_u$ the particle interface will become unstable. Therefore, the maximum cooling rate that maintains stable particle growth must be when the two times are equal:

$$\Delta t_0 = \Delta t_u \qquad (20)$$

Expressions for $\Delta t_0$ and $\Delta t_u$ will now be developed.

The time it takes for solute fields surrounding identical neighboring particles to interact is proportional to the square of half the distance separating the particles divided by the solute diffusivity in the liquid, or

$$\Delta t_0 = a \frac{(\Delta R)^2}{D_L} \tag{21}$$

where ΔR is half the distance separating a particle from its neighbor, and "a" is the pre-factor of the proportionality, which is of order unity.

To obtain an expression for $\Delta t_u$, standard expressions for the morphological stability of a flat solid-liquid interface were employed.[4] The amplification rate, ω, of a linear perturbation of wave-vector $K = \frac{2\pi}{\lambda}$ on an initially flat solid-liquid interface is:

$$\omega = VK - \frac{D_L \Gamma}{m_L C_0 (k-1)} K^3 \tag{22}$$

where λ is the wavelength of the perturbation. By differentiating Equation (22), setting $\frac{d\omega}{dK} = 0$ and V = V$_{max}$, the wave-vector for the perturbation being amplified the fastest, K$_{max}$, can be obtained:

$$K_{max} = \left[ \frac{V_{max} m_L C_0 (k-1)}{3 D_L \Gamma} \right]^{1/2} \tag{23}$$

where $V_{max}$ is the velocity of the fastest growing perturbation. Substituting Equation (23) into (22) gives the maximum amplification rate for the perturbation:

$$\omega_{max} = \frac{2}{3} \left[ \frac{(V_{max})^3 m_L C_0 (k-1)}{3 D_L \Gamma} \right]^{1/2} \tag{24}$$

The time required for the initially flat interface to become morphologically unstable, $\Delta t_u$, must be inversely proportional to the amplification rate of the fastest growing perturbation. This results in the expression

$$\Delta t_u = B\left(\frac{3D_L \Gamma}{(V_{max})^3 m_L C_0 (k-1)}\right)^{1/2} \tag{25}$$

where B is a pre-factor for the proportionality.

Now that experessions for both $\Delta t_0$ and $\Delta t_u$ have been developed, the final step is to express the maximum growth velocity in terms of the bulk cooling rate of the slurry. This is achieved with the use of Equation (19) developed in the previous section. As has been shown, the maximum growth velocity is reached just before the solute fields of neighboring particles begin to interact, which is precisely at the time $\Delta t_0$ given in Equation (21). Substituting Equation (21) into Equation (19) gives the maximum particle growth velocity

$$V_{max} = b'\left[\frac{\dot{T}_{max}}{|m_L|C_0(1-k)}\right](\Delta R) \tag{26}$$

where $b'$ is the new pre-factor combining the pre-factors a and A, and $\dot{T}_{max}$ is the maximum cooling rate that will allow stable growth. Substituting Equations (21), (25), and (26), into (20), and solving for ΔR yields:

$$\Delta R = C\left[\frac{(D_L)^3 \Gamma [m_L C_0(k-1)]^2}{(\dot{T}_{max})^3}\right]^{1/7} \tag{27}$$

where C is a pre-factor that combines the prefactors of the previous equations. It is expected to be of order unity by virtue of the fact that all these prefactors are also expected to be of order unity.

If $R_i$ is defined as the initial radius of the particle when the cooling rate is increased, and $R_T$ is again defined as the maximum radius the particle can grow to before impinging on neighboring particles, substituting ($R_T$-$R_i$) for $\Delta R$ and solving for $R_i$ gives the expression

$$R_i = R_T - C \left[ \frac{(D_L)^3 \, \Gamma \, [m_L C_0 (k-1)]^2}{(\dot{T}_{max})^3} \right]^{1/7} \qquad (28)$$

Dividing both sides of the equation by $R_T$ gives

$$\left(\frac{R_i}{R_T}\right) = 1 - \frac{C}{R_T} \left[ \frac{(D_L)^3 \, \Gamma \, [m_L C_0 (k-1)]^2}{(\dot{T}_{max})^3} \right]^{1/7} \qquad (29)$$

The solid fraction in a solidifying metal slurry, $f_s$, can be estimated by the expresson

$$f_s = \left(\frac{R_i}{R_T}\right)^3 \qquad (30)$$

Combining (29) and (30) gives a relationship between the solid fraction and the maximum cooling rate for interface stability:

$$f_s = \left[ 1 - \frac{C}{R_T} \left[ \frac{(D_L)^3 \, \Gamma \, [m_L C_0 (k-1)]^2}{(\dot{T}_{max})^3} \right]^{1/7} \right]^3 \qquad (31)$$

Alternatively, the relationship may be written in terms of grain density in the slurry, N, which is the number of particles per unit volume

$$N = \frac{1}{\frac{4}{3}\pi(R_T)^3} \tag{32}$$

yielding the final result

$$f_s = \left[1 - C\left(\frac{4\pi N}{3}\right)^{1/3}\left(\frac{(D_L)^3\,\Gamma\,[m_L C_0(k-1)]^2}{(\dot{T}_{max})^3}\right)^{1/7}\right]^3 \tag{33}$$

Figure 8 plots Equation (33) as fraction solid versus cooling rate for Al-4.5wt%Cu alloy for several values of particle density, assuming a prefactor C=1.2. As noted earlier, the prefactor is expected to be on the order of unity, and the value of 1.2 is selected here because of agreement with experiment, to be discussed in a later section of this paper. Each curve is a stability boundary. The region to the left of each curve represents combinations of solid fraction and cooling rate that will maintain particle stability, and the region to the right of the curve represents combinations leading to unstable particle growth. Stability is favored by low cooling rate, high fraction solid and high particle density. Figure 9 plots the particle radius versus cooling rate for different values of particle density by combining Equations (31) and (32). For a given grain density, stability is seen to be favored by a large initial particle radius.

Figures 8 and 9 now may be used to determine casting parameters needed to assure formation of a fully spheroidal microstructure at, or after, casting. This is done by considering also the ripening relationship, which may be written, for Al-4.5wt%Cu alloy[2] as:

$$R \approx 5t^{1/3} \tag{34}$$

where t represents the time in seconds in the liquid-solid temperature zone during heating, cooling, or isothermal holding, and R is in micrometers. The method of determining casting parameters is illustrated by two examples given in the Discussion section to follow.

## Experiments Performed

In preliminary experiments to test the model, a slurry of semi-solid Al-4.5% Cu alloy was produced and then rapidly solidified by drawing the slurry into a thin copper mold as described in previous literature.[2],[6] Measured solute content within the actual particles was then compared with that predicted by the LDC model. Figure 10(a) shows a particle from a specimen that was cooled for 20 seconds to just below the liquidus, while being vigorously stirred, and then quenched. During this 20 seconds the bulk melt cooling rate was measured to be 0.28 °C s$^{-1}$. The quench rate during solidification in the copper mold was estimated to be 375 °C s$^{-1}$. Figure 10(b) shows the solute concentration profile across the diameter of the particle. During the initial slow cooling, solute fields of the growing particles almost completely overlapped, and so solute distribution in the solid during this period is as would be predicted by the Scheil equation. Then there is an abrupt increase in solute content beginning at a radial distance of approximately 20μm, the radius at which the quench was initiated. This abrupt increase is clear qualitative evidence of solute build-up in the liquid, since, in the absence of such buildup, solute increase from microsegregation according to the Scheil equation would be gradual and small until well past the 30 μm radius.

Quantitative agreement of the model calculation with this experiment is seen by the close match of the experimental data with calculated curves in Figures 2, 3, and 10, all based on the LDC model. From the experimental data in Figure 10 it is seen that solid solute content in the particle at a radius of 40 µm is about 0.9 wt.% above that at radii within the 20µm original spheroid. Dividing this by k yields an increase of 4.7wt%Cu in the liquid, close to that given in Figure 2 for r = 40 µm. Multiplying the solute increase in the liquid by $m_L$ to get the increase in liquid content at the interface yields an undercooling of 16°C, close to the calculated value in Figure 3 of about 18°C. Finally, as shown in Figure 10(c), the calculated curve in Figure 10 matches well the experimental data. The solute content predicted by the Scheil Equation is also given for comparison with the LDC model and experimental data.

In a second series of experiments, small samples of fully-solidified Al-4.5wt%Cu alloy were cut from a small ingot produced with spheroidal microstructure. This initial structure is shown in Figure 11.

Each cylindrical sample was roughly 1 cm in diameter, and 1.5 cm in height. A small hole, slightly larger than 1 cm in diameter and 1 cm deep was drilled in the top of each sample, along the axis of the cylinder. A K-type thermocouple (0.1 cm dia.) was inserted in the hole to measure sample temperature. Samples were placed on a holder and inserted into a preheated furnace. They were reheated to three temperatures which corresponded to solid fractions of 0.25, 0.45, and 0.63 (approximated using the Scheil relation). For samples reheated to a solid fraction of 0.25, the metal was too fluid to hold its shape. These samples were held together during reheating by a thin-walled steel sleeve coated with boron nitride spray. The time to heat and equilibrate each samples at one of the three temperatures was always between 9 and 12 minutes.

After equilibrating at one of the three temperatures, the samples were cooled at various rates, ranging between 1 and 50 °C s$^{-1}$. The Biot number for the samples was approximated to be about 0.08; thus the entire sample cooled uniformly. After each sample had solidified completely, it was polished and etched with Keller's reagent. Between 5 and 10 micrographs were taken from each sample. The stability of the particles in each sample was assessed qualitatively by looking at the final morphology of what had been the solid-liquid interface.

As an example of the qualitative assessment of particle stability, Figures 12 (a)-(e) show the microstructure of samples reheated to a fraction solid of 0.25 and cooled at 1.1, 2.8, 4.2, 9, and 38 °C s$^{-1}$. The microstructures in Figures 12 (a) show stable particle interfaces which resulted from cooling the samples at 1.1°C s$^{-1}$, and all other cooling rates produced unstable particle interfaces.

Figure 13 is a summary of all the interface stability experiments conducted. The y-axis of the graph is solid fraction, and the x-axis is cooling rate, plotted on a logarithmic scale. The solid fraction-cooling rate combinations which produced samples with mostly stable or unstable particle interfaces are labeled as either circles or triangles, respectively. A clear trend line separating stable and unstable combinations of solid fraction and cooling rate is evident in Figure 13.

To compare experimental results with theory, $R_T$ was estimated from the micrographs of experimental samples (some of which are shown in Figure 12(a)-(e)). The average distance separating particle centers in the micrographs was estimated to be about 300 μm, yielding an $R_T$ of 150 μm. This value for $R_T$ was converted to a particle density estimate of 70 particles mm$^{-3}$ by using Equation (32). These experimental data inserted in Equation 32, with a pre-factor "C" of 1.2 result in exactly the curve separating the stable from the unstable points in Figure 13. This value of the pre-factor C is consistent with the expectation that it is near unity, as stated previously.

## Discussion

A solidification model has been presented for alloy solidification assuming spheroidal morphology of growing grains. The model considers the effect of limited diffusion of solute in the liquid, and can be used to calculate concentration gradients and undercoolings in solidifying alloys. The model is first solved numerically. At low cooling rates and high grain density, solute fields around solid particles overlap extensively and solidification obeys the well known Scheil relation. However, in special cases, substantial solute gradients and hence undercoolings result, and if sufficiently great, the spheroidal morphology breaks down, forming dendrites. For example, in one calculated example, the cooling rate of a slurry of particles 20 μm in diameter was suddenly increased from 0.28 °C s$^{-1}$ to 375 °C s$^{-1}$, at which time calculated undercooling in the liquid rose rapidly to a maximum of 18 °C, and then gradually decreased due to solute overlap. In this calculation a "volume element" of radius 50 μm was assumed ($R_T$=50 μm), corresponding to a grain density of 1900 particles mm$^{-3}$.

Of considerable practical importance for semi-solid metal processing is determining the maximum rate at which a slurry of known grain density, fraction solid and particle diameter can be cooled. This is done in this paper by first obtaining an approximate analytical solution to the numerical model above, and then combining that with stability theory. Following are two examples:

Consider first a semi-solid (thixocast) alloy billet reheated to a temperature where it is 50% liquid. The total time during casting and reheating that the billet is in the liquid-solid zone is 25 minutes, so the grain radius, $R_i$, at our starting point is approximately 120 μm, an estimate made using the ripening relation given by Equation (34). From Equations (30) and (32) the grain density is approximated to be roughly 70 grains mm$^{-3}$. Now, from either Figure 8 or 9, it is seen that cooling rate from the reheating temperature and before entering the die can be as high as about 6 °C s$^{-1}$ while maintaining the spherical morphology desirable for homogeneous flow and formability. Of course, more rapid cooling during solidification in the die might well result in a dendritic periphery of the otherwise spherical grains, resulting in much poorer formability.

Consider now cooling the alloy from the melt with vigorous agitation and rapid initial heat extraction to about 6% solid and then holding for the short time of 8 seconds. The cooling and agitation results in copious nucleation of small spheroids or perhaps nascent dendrites. The 8 second holding time ripens the grains to spheroids (if they were not already spherical) with a particle radius of approximately 10 μm, Equation (34). From Equations (30) and (32), the grain density is now approximated to be about 15,000 grains mm$^{-3}$. From either Figure 8 or 9 it is seen that cooling during processing and before entering the die can be as high as about 25 °C s$^{-1}$ while maintaining the desired spherical morphology for formability.

## Conclusions

1. Spheroidal growth is favored in semi-solid alloys by high grain density, high fraction solid and low cooling rate.

2. An analytic model for stability of spheroidal growth is presented and shown to agree with experiments on Al-4.5 wt. %Cu alloy.

3. The grain size and hence the grain density of a rheocast alloy can be estimated by the ripening relation and knowledge of the solidification time. Then, with this initial structure and known temperature, the stability model developed permits direct calculation of the maximum cooling rate that may be employed in subsequent processing while still maintaining the spheroidal structure.

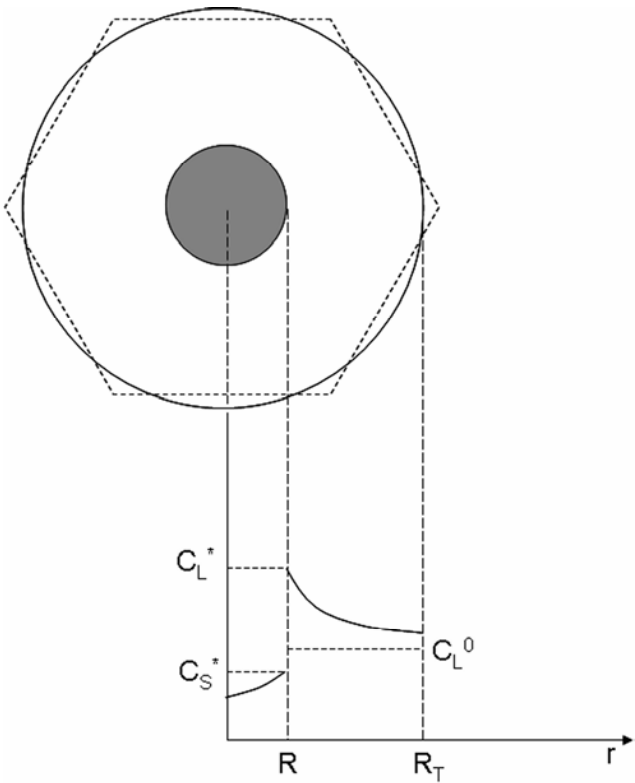

Figure 1. Schematic of a solidifying particle in rheocast alloy according to the liquid-diffusion-controlled (LDC) solidification model. The dashed hexagon represents the size of the actual grain after solidification is complete. The sphere of radius $R_T$ is used in the model to approximate this size.

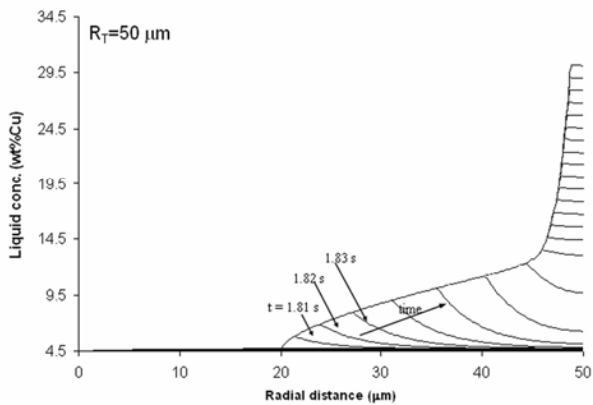

Figure 2. The LDC model of liquid composition as a function of radial distance with slow cooling (0.28 °C s$^{-1}$) for growth up to 20 μm particle radius, followed by fast cooling (375 °C s$^{-1}$). When the particle reaches 40 μm in radius, the composition in the liquid at radius $R_T$ (50 μm) begins to rise, marking the onset of solute field overlap.

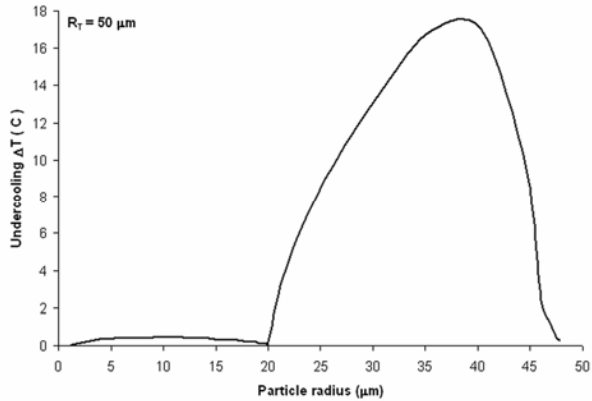

Figure 3. Undercooling in the liquid during solidification with slow cooling (0.28 °C s$^{-1}$) for growth up to 20 μm particle radius, followed by fast cooling (375 °C s$^{-1}$). When the particle reaches 40 μm in radius, the undercooling begins to diminish due to a rise in the solute concentration in the liquid at radius R$_T$ (50 μm), marking the onset of solute field overlap.

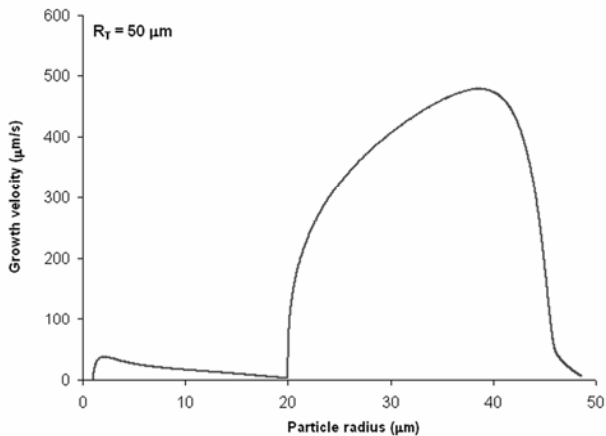

Figure 4. The LDC model prediction of growth velocity as a function of particle radius for the case of slow cooling (0.28 °C s$^{-1}$) for growth up to 20 μm particle radius, followed by fast cooling (375 °C s$^{-1}$). The growth velocity is shown to decrease suddenly due to solute field overlap when the particle reaches 40 μm in radius.

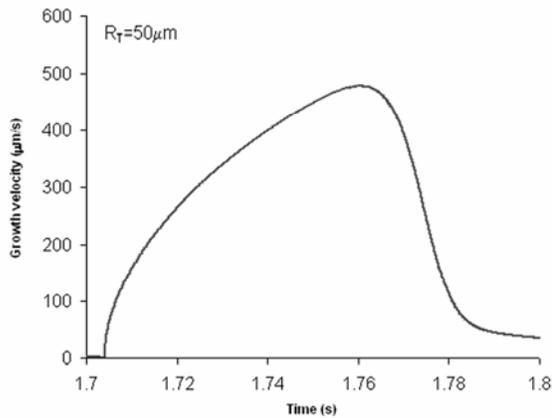

Figure 5. The LDC model prediction of growth velocity as a function of time for the case of slow cooling (0.28 °C s$^{-1}$) for growth up to 20 μm particle radius, followed by fast cooling (375 °C s$^{-1}$). The growth velocity is shown to decrease suddenly, marking the time when solute field overlap begins.

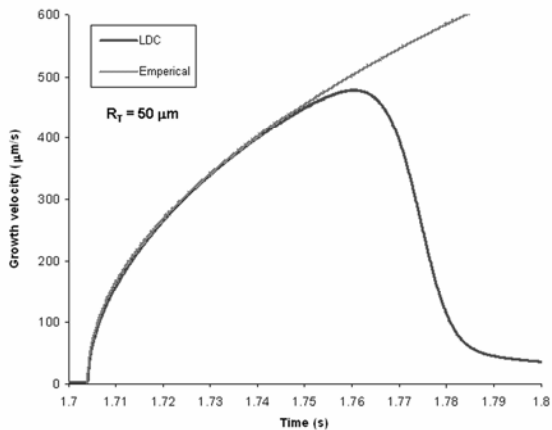

Figure 6. Comparison of the derived relationship between particle growth velocity and cooling rate with the LDC model prediction for the case of slow cooling (0.28 °C s$^{-1}$) for growth up to 20 μm particle radius, followed by fast cooling (375 °C s$^{-1}$). The derived expression follows the LDC growth velocity curve very well up until the maximum growth velocity, which occurs when solute field overlap begins.

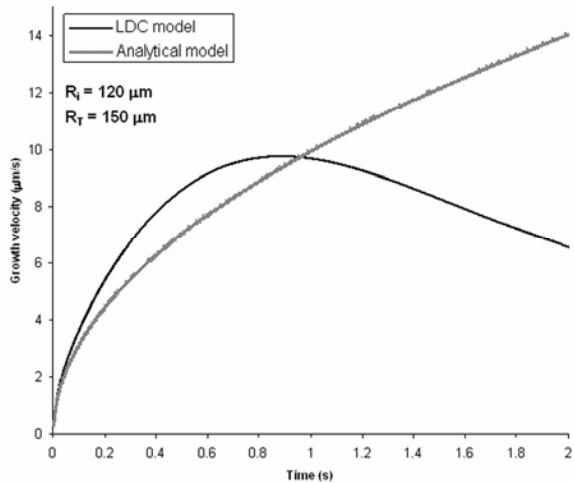

Figure 7. Comparison of the derived relationship between particle growth velocity and cooling rate with the LDC model prediction for the case of cooling at 2 °C s$^{-1}$. The initial particle radius is 120 μm, growing to a final particle radius of 150 μm. The derived expression slightly underestimates the growth velocity but again approximates the growth of the particle until the time when solute field overlap begins.

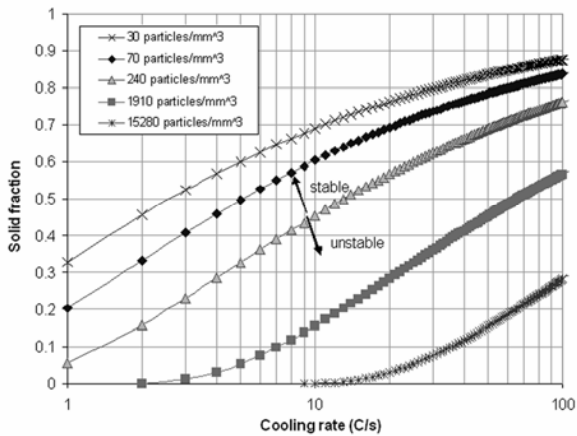

Figure 8. Particle stability model for rheocast Al-4.5wt%Cu alloy which considers solute field overlap during solidification. Curves for slurries with varying particle densities are given. For a given particle density, the region to the left of the curve represents combinations of solid fraction and cooling rate that will lead to stable particle growth, and the region to the right of the curve represents combinations that will lead to unstable particle growth. All curves were produced using a pre-factor of C=1.2.

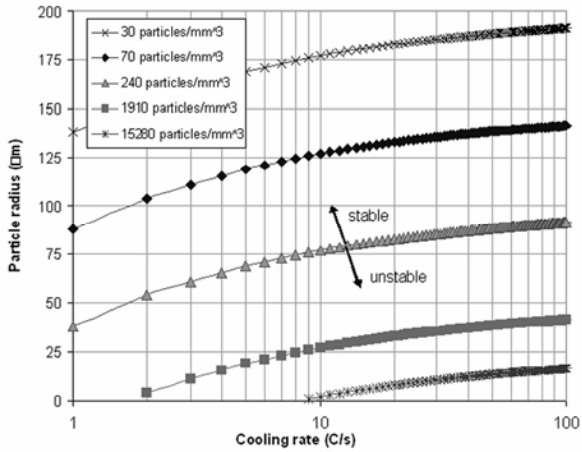

Figure 9. Relationship between cooling rate and minimum particle radius for rheocast Al-4.5wt%Cu alloy according to the stability model which considers solute field overlap during solidification. For a given particle density, the region to the left of the curve represents particle radii that will be stable at corresponding cooling rates, and the region to the right of the curve represents combinations that will lead to unstable particle growth. All curves were produced using a pre-factor of C=1.2.

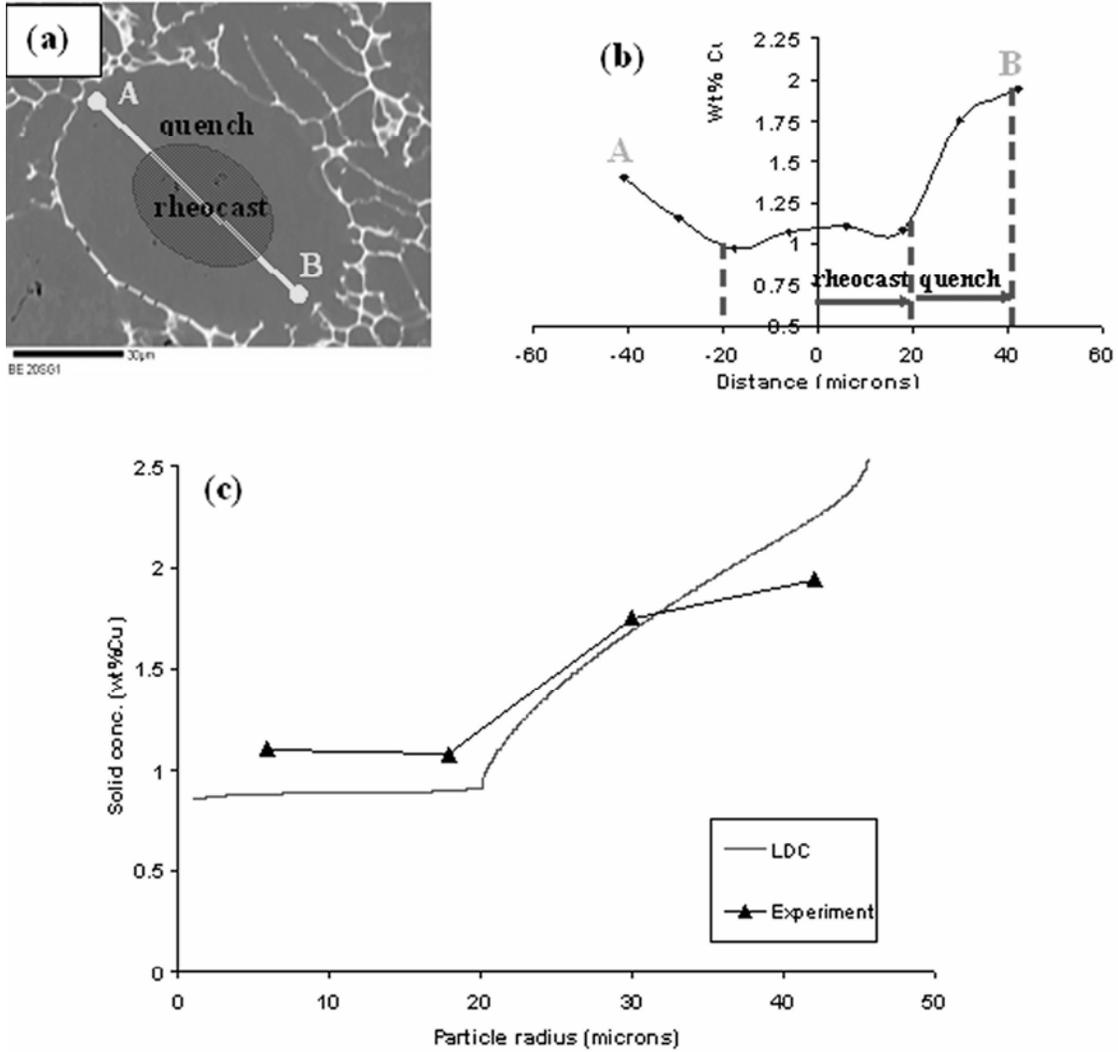

Figure 10. (a) Back-scattered electron image of a particle in rheocast Al-4.5wt%Cu alloy (b) the copper content within the particle measured by a microprobe line-scan (c) Comparison of the microprobe data with the predicted values given by the LDC model with slow cooling (0.28 °C s$^{-1}$) for growth up to 20 μm particle radius, followed by fast cooling (375 °C s$^{-1}$).

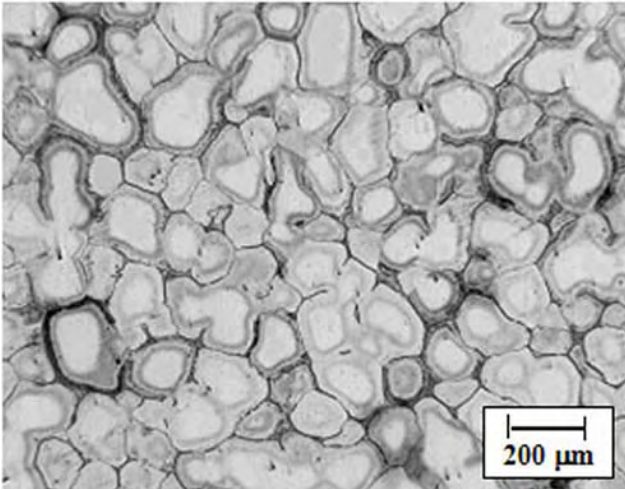
Figure 11. Rheocast Al-4.5wt%Cu showing the non-dendritic, spheroidal solid morphology.

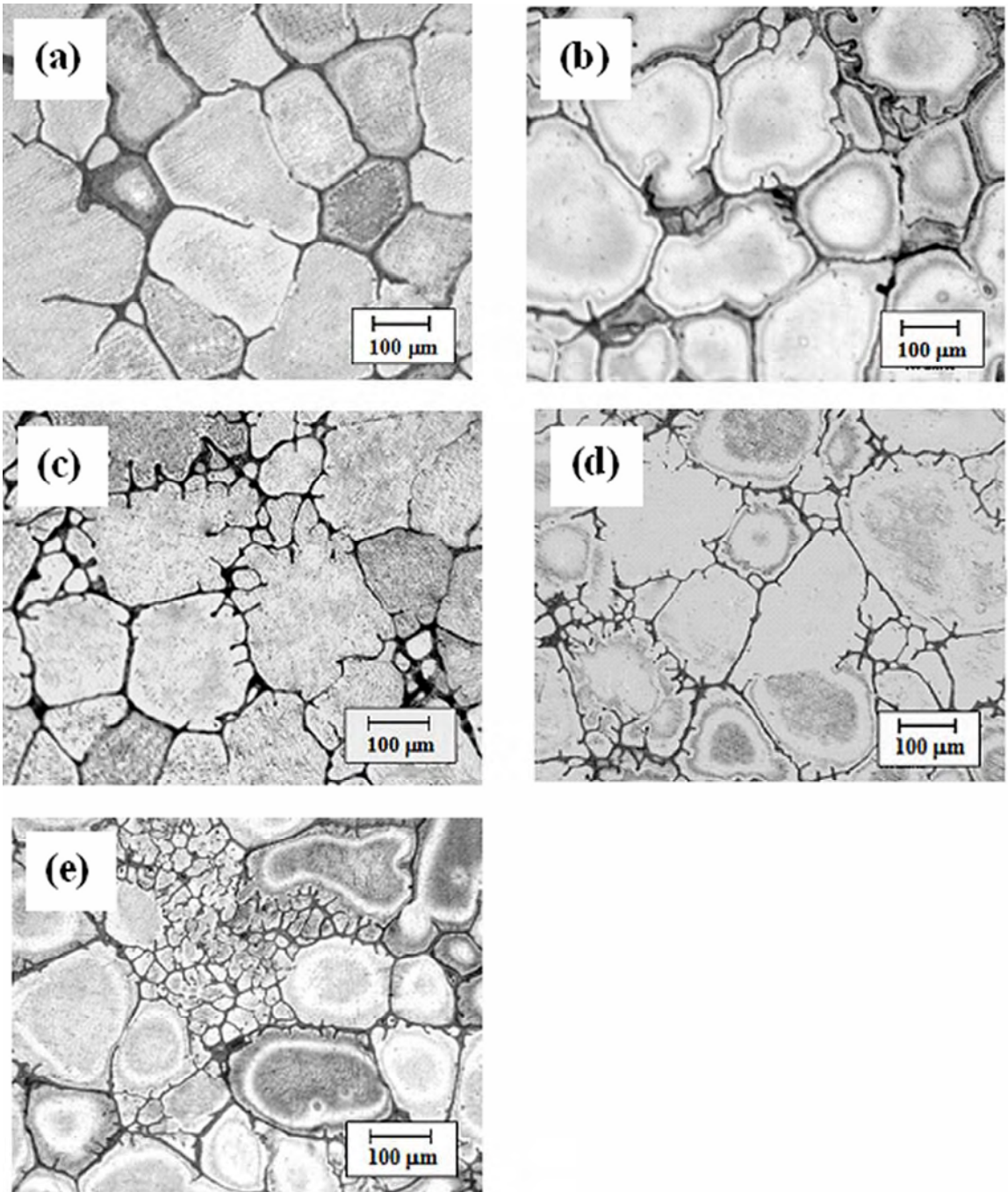

Figure 12. Microstructures of rheocast Al-4.5wt%Cu alloy reheated to a solid fraction of 0.25 and then cooled at (a) 1.1 (b) 2.8 (c) 4.2 (d) 9 (e) 38 °C s$^{-1}$.

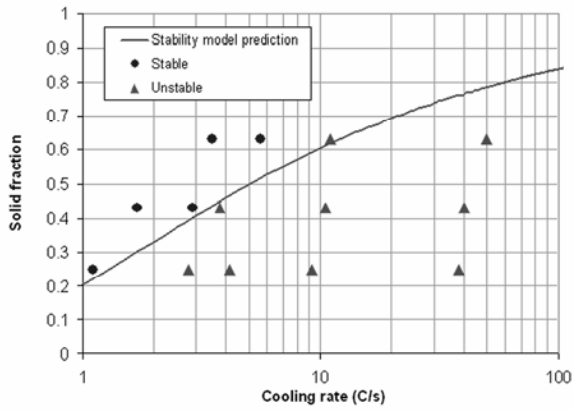

Figure 13. Comparison of the stability model prediction with experimental data for rheocast Al-4.5wt%Cu alloy. A particle density of 70 particles mm$^{-3}$ was obtained from experimental data and inputted into the stability model along with the pre-factor of C=1.2.